\documentclass{ws-procs9x6-cpt22}
\begin{document}

\newcommand{\refeq}[1]{(\ref{#1})}
\def\etal {{\it et al.}}

\def\nr{\text{NR}}
\def\nrtemplate#1#2#3{#1^{\nr#3}_{#2}}
\def\anr#1{\nrtemplate{a}{#1}{}}
\def\cnr#1{\nrtemplate{c}{#1}{}}
\def\gzBnr#1{\nrtemplate{g}{#1}{(0B)}}
\def\goBnr#1{\nrtemplate{g}{#1}{(1B)}}
\def\HzBnr#1{\nrtemplate{H}{#1}{(0B)}}
\def\HoBnr#1{\nrtemplate{H}{#1}{(1B)}}

\def\K{\mathcal K}

\title{Lorentz Violation, CPT violation, and Spectroscopy Experiments}

\author{Arnaldo J.\ Vargas}

\address{Laboratory of Theoretical Physics Department of Physics, University of Puerto Rico, 
Río Piedras 00936, Puerto Rico}

\begin{abstract}
This presentation discusses some of the signals for Lorentz violation potentially observable in atomic spectroscopy and clock-comparison experiments. The emphasis of the discussion is on how the angular-momentum quantum numbers of the states involved in the transition determine the properties of the possible sidereal variation of the transition frequency. 

\end{abstract}

\bodymatter

\section{Motivation and introduction}

The Standard-Model Extension (SME) is an effective field theory that contains all the possible Lorentz-violating deviations from the Standard Model and General Relativity.\cite{SME} A reason for the creation of the SME is to facilitate the systematic search for Lorentz and CPT violations by classifying all Lorentz-violating operators in terms of measurable SME coefficients. We can quantify the reliability of Lorentz symmetry by measuring how consistent with zero are all of the SME coefficients. 

The first models for searching for Lorentz violation in atomic experiments, called in this presentation atomic SME models, were limited to Lorentz-violating operators of mass dimensions three and four called minimal operators.\cite{Prev} Lorentz-violating operators of mass dimensions five or higher are called nonminimal Lorentz-violating operators. The minimal atomic SME models, the models limited to minimal operators, were used to identify signals for Lorentz violation that were targeted by clock comparison experiments with ordinary matter and exotic atoms.\cite{AtomSpec} Some of the historical and current best bounds on many SME coefficients are the results of these experimental studies.\cite{datatables}

Any systematic search for Lorentz violation must consider minimal and nonminimal Lorentz-violating operators. The limitation on the first SME models to consider only minimal Lorentz-violating operators was for practical reasons. For example, it took more than a decade after the creation of the SME for all the fermion Lorentz-violating operators of arbitrary mass dimensions to be classified.\cite{KoMe13} After the theoretical advancements in recent years, models for testing Lorentz symmetry with atomic experiments that include nonminimal operators have been obtained.\cite{GoKo14, KoVa15, KoVa18}

This presentation is based on SME nonminimal atomic models introduced in three publications. The first publication considers atomic spectroscopy experiments with muonium and muonic atoms.\cite{GoKo14} The second publication\cite{KoVa15} considers two-fermion atoms such as hydrogen, antihydrogen, and positronium as well as deuterium and some simple molecules. The third publication extends the analyses to multiple fermion atoms with applications to microwave and optical clocks.

\section{NR coefficients and atomic energy levels}

The nonrelativistic (NR) coefficients $\anr{kjm}$, $\cnr{kjm}$, $\HzBnr{kjm}$, $\HoBnr{kjm}$, $\gzBnr{kjm}$, and $\goBnr{kjm}$, which are defined in Ref.~\refcite{KoMe13}, are the effective SME coefficients used in the models discussed in this presentation.\cite{GoKo14, KoVa15, KoVa18} The Lorentz-violating operators associated with the NR coefficients are expressed as the components of spherical tensors.\cite{KoMe13} The component of a spherical tensor is an object that rotates as the angular momentum eigenstates $|jm\rangle$. For the states $|jm\rangle$, $j$ is the rank of the spherical tensor and $m$ is the component. Following a similar notation, the $j$ index of the NR coefficient specifies the rank of the spherical tensor $T^{(j)}_m$ associated with the coefficient, and the $m$ index specifies the component. The meaning of the other indices of the NR coefficients are explained in Ref.~\refcite{KoMe13}.

The leading-order Lorentz-violating atomic energy shift is obtained from the expectation value of the Lorentz-violating perturbation. The perturbation  can be expressed as\cite{GoKo14, KoVa15, KoVa18}
\begin{equation}
\delta h=\sum_j\sum^j_{m=-j} \K_{jm} T^{(j)}_m,
\end{equation}
where $\K_{jm}$ represents a generic NR coefficient and $T^{(j)}_m$ represents a generic Lorentz-violating operator expressed as the component of a spherical tensor. Representing the atomic state as $|F\rangle$, where we are supressing all the quantum numbers except the total angular momentum quantum number $F$, the leading-order energy shift is given by\cite{GoKo14, KoVa15, KoVa18}
\begin{equation}
\delta \epsilon=\sum_j\sum^j_{m=-j} \K_{jm}\langle F| T^{(j)}_m|F\rangle.\label{eshift}
\end{equation}

An advantage of expressing the Lorentz-violating operators in terms of spherical tensors is that we can use the Wigner-Eckart theorem\cite{WiEc}. This theorem implies that $\langle F| T^{(j)}_m|F\rangle=0$ if $j>2F$ and therefore, only the NR coefficients with index $j\leq 2F$ can contribute to the Lorentz-violating shift for an atomic state with quantum number $F$. In the case of the $nS_{1/2}$ atomic state of hydrogen, the maximum possible value for $F$ is $F=1$, and therefore, only NR coefficients with $j\leq 2$ contribute to the energy shift. Evidently, it is impossible to measure all the NR coefficients by only studying transitions between $nS_{1/2}$ states as the NR coefficients with $j>2$ do not contribute to the energy of these states. The situation is different in the minimal SME as all coefficients in the minimal SME atomic models contribute to the energy of the $nS_{1/2}$ states. Therefore, in the context of the minimal SME is enough to study transitions between $nS_{1/2}$ states. In the nonminimal SME models, transitions between states with higher values of $F$ are necessary to measure all the NR coefficients. For instance, only transitions involving states with $F\geq2$ are sensitive to NR coefficients with $j=4$. 

\section{NR coefficients and harmonics of the sidereal frequency} 

A possible signal for Lorentz violation is if the atomic transition frequency depends on the orientation of the atom. We can rotate the atom by fixing the atomic state relative to the quantization axis defined by an external uniform magnetic field. In this scenario, we can rotate the atomic states by rotating the magnetic field. The most common situation is when the orientation of the magnetic field is fixed relative to a laboratory on the surface of the Earth. In this case, the magnetic field and the atomic state rotates with the Earth, and the signal for Lorentz violation is a sidereal variation of the transition frequency.\cite{GoKo14, KoVa15, KoVa18} For now, we will consider only the variations in the transition frequency due to the Earth's rotation and ignore the effect of the motion of the Earth around the Sun.  

In general, the sidereal variation of the Lorentz-violating shift to the transition frequency can be expressed as
\begin{equation}
\delta \nu=A_{0}+\sum_{q}(A_q\cos{(q\omega_\oplus T)}+B_q \sin{(q\omega_\oplus T)}) \label{sid},
\end{equation}
where $\omega_\oplus\simeq 2\pi/({\rm 23h\,56 m})$ is the sidereal frequency, $T$ is the time in the Sun-centered frame\cite{datatables} and $q\geq 1$. The NR coefficients that contribute to the amplitude $A_q$ and $B_q$ of the hamonics of the sidereal frequency are determined by the abosulute value $|m|$ of the $m$-index. If $q\ne|m|$, then the NR coefficient $\K_{jm}$ cannot contribute to the amplitudes $A_q$ or $B_q$.\cite{KoVa15, KoVa18} For example, the coefficient $\HoBnr{432}$ has $m=2$ and therefore, it can only contribute to $A_2$, $B_2$, or both.     Consequently, the goal of any sidereal variation study of the transition frequency $\nu$ is to measure the amplitudes of the harmonics of the sidereal frequency that contribute to $\delta \nu$. 

The highest harmonic of the sidereal frequency that contributes to $\delta \nu$ is determined by the angular momentum quantum numbers of the states involved in the transition.\cite{KoVa15, KoVa18} It is worth contrasting this to the minimal SME case where the highest harmonic that contributes to $\delta \nu$ is the first harmonic of the sidereal frequency in the absence of boost corrections.  Assuming that $F_{\rm max}$ is the highest value of $F$ for the states involved in the transition, the highest harmonic of the sidereal frequency capable of contributing to $\delta \nu$ is $q=2F_{\rm max}$.  As an illustration, if $F_{\rm max}=3$, the highest harmonic of the sidereal frequency that can contribute to $\delta \nu$ is the sixth harmonic.  

Based on the previous discussion, we might conclude that the transition frequency between atomic states with $F=0$ has no sidereal variation. In general, this is false. As mentioned, we have been ignoring the contribution from the boost transformation between the lab and the Sun-centered frame. We justify this because of the slow speed, around  $10^{-4}$, between the frames. However, even the transition frequency between atomic states with $F=0$ can have a sidereal variation if we include the boost corrections. Additionally, the transition frequency can gain an annual variation due to the Earth's orbital motion.\cite{KoVa15, KoVa18} The aforementioned annual variation is important as it is the dominant signal for Lorentz violation for many potential studies of Lorentz violation with optical clocks.\cite{KoVa18}

\section{NR coefficients and antihydrogen} 

 CPT violation implies Lorentz violation in any realistic interacting field theory.\cite{Gr02} Hence, the SME can be used to facilitate the systematic search for CPT violation by classifying all the CPT-violating terms. In the models discussed in this presentation, the operators associated with the $a$-type and $g$-type NR coefficients are CPT-violating operators. A signal for CPT violation produced by these CPT-violating operators is a discrepancy between the spectrum of hydrogen and antihydrogen.\cite{KoVa15} Several groups are measuring or planning to measure antihydrogen transition frequencies to compare them to their hydrogen counterpart.\cite{Alpha, ASACUSA}  All the antihydrogen transitions considered by these collaborations are between states with $F\leq 1$. Based on the previous discussion, these measurements will be sensitive to a small subset of the CPT-violating operators in our models. Therefore, any systematic search of CPT violations must measure antihydrogen transitions involving a state with $F\geq 1$.  

\section*{Acknowledgments}
This work was supported 
by Department of Energy grant {DE}-SC0010120
and by the Indiana University Center for Spacetime Symmetries.

\end{document}